\documentstyle[prb,aps,psfig,twocolumn]{revtex}
\def\sech{\mbox{sech}}

\begin{document}
\title{
Coupled nonlinear Schr\"{o}dinger field equations
for electromagnetic wave propagation in nonlinear left-handed 
materials
}
\author{N. Lazarides$\ ^{1,2}$ and G. P. Tsironis$\ ^{3}$
}
\address{
$\ ^{1}$Department of Materials Science and Technology,
  University of Crete, P. O. Box 2208, 71003,
  Heraklion,  Greece \\
$\ ^{2}$Department of Natural Sciences,
  Technological Education Institute of Crete,
  P. O. Box 1939, 71004, Heraklion, Greece \\
$\ ^{3}$Department of Physics, University of Crete,
  and FORTH,
  P. O. Box 2208, Heraklion, 71003, Greece
}

\vspace{-0.9cm}
\maketitle
\begin{abstract}
For an isotropic and homogeneous nonlinear left-handed materials,
for which the effective medium approximation is valid,
Maxwell's equations for electric and magnetic fields lead naturally,
within the slowly varying envelope approximation,
to a system of coupled nonlinear Schr\"{o}dinger equations.
This system is equivalent to the well-known Manakov model that
under certain conditions, is completely integrable and admits bright 
and dark soliton solutions.
It is demonstrated that left- and right-handed (normal) nonlinear 
mediums may have compound bright and dark soliton solutions, respectively.
These results are also supported by numerical calculations. 

\end{abstract}


\vspace{-0.9cm}

Recently the study of the electromagnetic (EM) properties
of artificial complex media with simultaneously negative 
dielectric permittivity $\epsilon_{eff}$ and magnetic 
permeability $\mu_{eff}$ has been the subject of great attention.
Such media are usually referred to as left-handed materials (LHMs),
\cite{veselago}
and they demonstrate a number of peculiar properties:
reversal of Snell's law of refraction,
reversal of the Doppler shift, counter-directed Cherenkov
radiation cone, negative refraction index, 
the refocusing of EM waves from a point sourse, etc.
The  above-mentioned properties follow directly from Maxwell's 
equations with appropriate constitutive relations.
Pendry\cite{pendry} has proposed the intriguing possibility
to exploit the negative refraction index property of the LHMs
in order to overcome known problems with common lenses to 
achieve a perfect lens that would focus both the propagating as 
well as the evanescent spectra.

Typical LHM are composed of a combination of a regular array of
electrically small resonant particles referred to as split-ring
resonators (SRR's) and a  regular array of conducting wires,
\cite{pendry,smith,shelby,pendry123,walser}
responsible for the negative $\mu_{eff}$ and $\epsilon_{eff}$,
respectively.
The size and spacing of the conducting elements of which the medium is
composed is assumed to be on a scale much smaller than the wavelengths 
in the frequency range of interest, 
so that  the composite medium may be considered
as a continuous and a homogeneous one (effective medium approximation).
Thus far, almost all properties of LHMs were studied  in the linear
regime of wave propagation, when both $\epsilon_{eff}$ and $\mu_{eff}$
are considered to be independent of the field intensities.
However, nonlinear effects in LHMs have been recently taken 
under consideration by some authors.\cite{zharov,obrien,lapine}
Zharov {\it et al},\cite{zharov}
considered a two-dimensional periodic structure created
by arrays of wires and  SRR's embedded into a nonlinear dielectric,
and they  calculated  $\epsilon_{eff}$ and $\mu_{eff}$
for a Kerr-type dielectric permitivity.
They showed that the magnetic field intensity couples to the 
magnetic resonance of the SRR in a nontrivial way, and that 
changing the material properties from left- to right-handed and back
is allowed by varying the field intensity.
The study of the nonlinear properties
of LHMs could facilitate future efforts in creating tunable structures
where the field intensity changes their transmission properties.

In the present work  we show that for an isotropic, homogeneous,
quasi-one-dimensional LHM, Maxwell's equations with  
nonlinear constitutive relations lead naturally
to a system of coupled nonlinear Schr\"{o}dinger (CNLS) equations
for the  envelopes of the propagating electric and magnetic fields.
This system is equivalent to the Manakov model\cite{manakov}
that under certain conditions, 
admits soliton solutions consisting of two components
(vector solitons). 
For specific parameter choices, corresponding to
either a left-handed or a right-handed medium, we find 
compound dark and bright soliton solutions, respectively. 
The constitutive relations can be generally written as
\begin{eqnarray}
\label{const1}
 {\bf D} &=& \epsilon_{eff} {\bf E} = \epsilon {\bf E} + {\bf P_{NL}} \\ 
\label{const2}
 {\bf B} &=& \mu_{eff} {\bf H} = \mu {\bf H} + {\bf M_{NL}} , 
\end{eqnarray}   
where  ${\bf E}$ and ${\bf H}$ are the electric and magnetic field 
intensities, respectively,  ${\bf D}$ is the electric flux density,
and ${\bf B}$ is the magnetic induction.
The linear dielectric and magnetic responses of the LHM are described 
by $\epsilon$ and $\mu$, respectively, while 
${\bf P}_{NL} = \epsilon_{NL} {\bf E}$ and ${\bf M}_{NL}  = \mu_{NL}  {\bf H}$
are the nonlinear electric polarization and the nonlinear magnetization
of the medium, respectively.

It is known that $\epsilon_{eff}$ and $\mu_{eff}$ in a LHM have 
to be dispersive, otherwise the energy density could be negative.
\cite{veselago}
Their frequency dispersion, including nonlinear effects
(but neglecting losses), is given by\cite{zharov} 
\begin{eqnarray}
  \label{eofomega}
    \epsilon_{eff} (\omega) &=& \epsilon_0 
       \left( \epsilon_D (|E|^2) - \frac{\omega_p^2}{\omega^2} \right) \\
  \label{mofomega}
   \mu_{eff} (\omega) &=& \mu_0 \left( 1 - \frac{ F \omega^2}
        { \omega^2 - \omega^2_{0NL} (|H|^2) } \right) ,
\end{eqnarray}	
where $\omega_p$ is the plasma frequency, $F$ is the filling factor,
$\omega_{0NL} = \omega_{0NL} (|H|^2)$ is the nonlinear resonant 
SRR frequency, and 
$\epsilon_D (|E|^2)= \epsilon_{D0} + \alpha |E|^2$,
with $\alpha$ the strength of the nonlinearity. Positive (negative)
$\alpha$ corresponds to a focusing (defocusing) dielectric.
For a linear dielectric
$\omega_{0NL}(|H|^2) \rightarrow \omega_0$,
where $\omega_0$ is the linear resonant SRR frequency.
Then Eqs. (\ref{eofomega}-\ref{mofomega}) reduce to previously
known expressions.\cite{pendry123,gorkunov}
The parameters $F$, $\omega_p$, and $\omega_0$ are related to geometrical
and material parameters of the LHM components.
Although $\epsilon_{eff}$ can be readily put in the form 
$\epsilon + \epsilon_{NL} ( |E|^2 )$, for $\mu_{eff}$ this is not
an obvious task, 
since  $\mu_{eff}=\mu_{eff} (\omega_{0NL})$, and $\omega_{0NL}$ depends on 
$|H|^2$ as\cite{zharov,obrien}
\begin{eqnarray}
\label{poly}
 \alpha \, \Omega^2 \, X^6 \, |{\bf H}|^2 
   = A^2 \,  E_c^2 \, (1 - X^2) (X^2 - \Omega^2)^2 ,
\end{eqnarray}
where $X= \omega_{0NL} / \omega_0$, $\Omega = \omega / \omega_0$,
$E_c$ is a characteristic (large) electric field,
and $A$ is a function of physical and geometrical parameters.
\cite{zharov,obrien}
The + (-) sign corresponds to a focusing (defocusing) dielectric.
Our $\alpha$ is related to the parameters of Eq. (\ref{poly})
as $\alpha=\pm1/E_c^2$.
Choosing $f_p= \omega_p /2\pi =10 ~GHz$ and $f_0= \omega_0 /2\pi =1.45 ~GHz$, 
left-handed behavior appears in a narrow frequency band 
(from $f=1.45~ GHz$ to $1.87~ GHz$). 
For relatively small fields when $\mu_{eff}$ is trully field dependent,
one may consider that the magnetic nonlinearity
$\mu_{eff} = \mu + \mu_{NL} (|H|^2)$ is of the Kerr type, i.e. 
$\mu_{NL} (|H|^2) = \beta\,|H|^2 $ (Fig. 1).
The strength of the magnetic nonlinearity $\beta$ can be treated as a 
fitting parameter.
\begin{figure}[h]
\centerline{\hbox{
  \psfig{figure=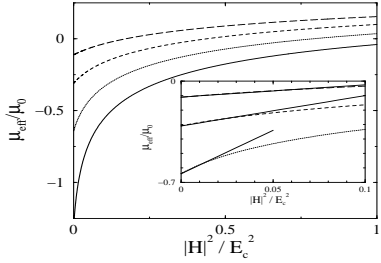,height=35mm,width=50mm,angle=-90}
}}
  \caption{
$\mu_{eff}$  as a function of $|H|^2 / E_c^2$,
for $\Omega=1.1$ (solid), $1.15$ (dotted), $1.20$
(dashed),  $1.25$ (long-dashed), $A=3$, $F=0.4$, and 
$\hat{\alpha}>0$. 
Inset: Fitting to a line of the three first curves of 
the main figure for relatively small fields.
}
\end{figure}

Using Eqs. (\ref{const1})-(\ref{const2}) and known identities, 
we get quite general vector wave equations for the fields ${\bf E}$ 
and ${\bf H}$ 
\begin{eqnarray}
\label{weq1}
  \nabla^2 {\bf E} - \mu \epsilon \frac{\partial^2 {\bf E}}{\partial t^2} 
  - \mu \frac{\partial^2 {\bf P}_{NL}}{\partial t^2} 
  -  \nabla (\nabla\cdot {\bf E} ) = \nonumber \\ 
 \frac{\partial}{\partial t} [ (\nabla \mu_{NL} ) \times {\bf H} ]
 + \frac{\partial}{\partial t} 
 \left[ \mu_{NL}  \frac{\partial}{\partial t}
 (\epsilon {\bf E} + {\bf P}_{NL} ) \right]  \\ 
\label{weq2}
  \nabla^2 {\bf H} - \mu \epsilon \frac{\partial^2 {\bf H}}{\partial t^2} 
    - \epsilon \frac{\partial^2 {\bf M_{NL}}}{\partial t^2} 
    -  \nabla (\nabla\cdot {\bf H} ) = \nonumber \\ 
 -\frac{\partial}{\partial t} [ (\nabla \epsilon_{NL} ) \times {\bf E} ]
 + \frac{\partial}{\partial t} 
   \left[ \epsilon_{NL}  \frac{\partial}{\partial t}
 (\mu {\bf H} +  {\bf M}_{NL}) \right] .
\end{eqnarray}
Consider an $x-$polarized plane wave  
with frequency $\omega$ propagating along the $z-$ axis:
\cite{hilesoerensen} 
\begin{eqnarray}
\label{tem}
 {\bf E} = ( E (z,t) , 0 ,0 ) , \qquad {\bf H} = (0, H (z,t), 0) ,
\end{eqnarray} 
where 
\begin{eqnarray}
  \label{monochromatic}
  E (z,t) = q (z,t) e^{i(kz -\omega t)} \,  \,  
  H (z,t) = p (z,t) e^{i(kz -\omega t)},
\end{eqnarray}
with  $k$ being the wavenumber.
The envelopes $q(z,t)$ and $p(z,t)$ of $E$ and $H$, respectively,
change slowly in $z$ and $t$. We therefore introduce the slow variables
\begin{eqnarray}
   \label{slow}
    \xi = \varepsilon ( z - \omega't)  \qquad \tau = \varepsilon^2 t ,
\end{eqnarray}
where $ \varepsilon$ is a small parameter, and 
$\omega' = \partial \omega / \partial k$ is the group velocity of the 
wave. 
Taking into account Eqs. (\ref{tem}-\ref{monochromatic}),
substituting slow variables into Eqs. (\ref{weq1}-\ref{weq2}), 
assuming that $\alpha = \hat{\alpha} \varepsilon^2$,
$\beta = \hat{\beta} \varepsilon^2$, 
and expressing $p$ and $q$ as an asymptotic 
expansion  in terms of $\varepsilon$
\cite{hilesoerensen}
\begin{eqnarray}
  \label{aexpansion}
    q(\xi,\tau) &=& q_0 (\xi,\tau) + \varepsilon q_1 (\xi,\tau) +
      \varepsilon^2  q_2 (\xi,\tau) + \cdots \nonumber \\ 
    p(\xi,\tau) &=& p_0 (\xi,\tau) + \varepsilon p_1 (\xi,\tau) +
      \varepsilon^2  p_2 (\xi,\tau) + \cdots , 
\end{eqnarray}
we get various equations in increasing powers of $\varepsilon$.
The leading order problem gives the dispersion relation $\omega = c k$,
where  $c = \sqrt{ 1 / \epsilon \mu }$.
At ${\cal O}(\varepsilon^1)$, the group velocity is given as
$\omega' = k c^2 / \omega$. At ${\cal O}(\varepsilon^2)$,
we obtain
\begin{eqnarray}
  \label{nls1}
   i  \frac{\partial q_0}{\partial \tau} 
   + \frac{\omega''}{2} \frac{\partial^2 q_0}{\partial \xi^2}
   + \frac{\omega c^2}{2} 
   (  \hat{\alpha} \mu |q_0|^2 + \epsilon \hat{\beta} |p_0|^2 ) q_0 &=& 0 \\ 
  \label{nls2}
  i  \frac{\partial p_0}{\partial \tau} 
 + \frac{\omega''}{2} \frac{\partial^2 p_0}{\partial \xi^2}
 + \frac{\omega c^2}{2} 
 ( \epsilon \hat{\beta} |p_0|^2 + \hat{\alpha} \mu |q_0|^2 ) p_0 &=& 0 ,
\end{eqnarray}
where $\omega'' = ( c^2 - \omega'^2 ) / \omega$, 
$\tau$ is the slow time,  and $\xi$ the slow space variable moving
at the linear group velocity. 
By rescaling $\tau$, $\xi$ and the amplitudes $q_0$, $p_0$
according to
\begin{eqnarray}
  \label{rescaling}
   \xi = X , \qquad T ={\omega'' \tau}/{2}, \\
   Q = \sqrt{\left| {\Lambda_q}/{\omega''} \right| } q_0, \qquad
   P = \sqrt{\left| {\Lambda_p}/{\omega''} \right| } p_0
\end{eqnarray}   
where $\Lambda_q = \omega c^2 \mu \hat{a}$ and 
$\Lambda_p = \omega c^2 \epsilon \hat{\beta}$, we get 
\begin{eqnarray}
   \label{eqs22a}
    i Q_T + Q_{XX} + \left( \sigma_q |Q|^2 + \sigma_p |P|^2 \right) Q =0 \\ 
   \label{eqs22b}
    i P_T + P_{XX} + \left( \sigma_p |P|^2 + \sigma_q |Q|^2 \right) P =0 , 
\end{eqnarray}
where $ \sigma_{q,p} \equiv sign(\Lambda_{q,p})$.  
Eqs. (\ref{eqs22a}-\ref{eqs22b}) is a special case of the fairly 
general and frequently studied system of CNLS equations 
(Manakov model) known to be completely integrable
for $ \sigma_{q} =  \sigma_{p} = \sigma$.\cite{manakov}
A number of bright and  dark soliton solutions have been obtained for
Eqs. (\ref{eqs22a}-\ref{eqs22b}) when $ \sigma=\pm 1$. 
\cite{park,yang,jakubowski,haelterman,christodoulides,kivsharsheppard}
There is evidence that single-soliton (single-hump) solutions
are stable while multi-hump are not.\cite{jakubowski}
The sign of the products $\mu \hat{\alpha}$ and $\epsilon \hat{\beta}$
determine the type of nonlinear self-modulation
(self-focusing or self-defocusing) effects which will occur.
For  $\sigma=\pm 1$
both fields experience the same type of nonlinearity.

For $\epsilon,\, \mu > 0$ and $\hat{a}, \hat{\beta} >0$
we have $\sigma=+1$ , 
and the system of Eqs. (\ref{eqs22a}-\ref{eqs22b})
accepts solutions of the form\cite{yang} 
\begin{eqnarray}
  \label{35}
    Q(X,T) = u(X) e^{i \nu^2_q T} \qquad
    P(X,T) = v(X) e^{i \nu^2_p T} ,
\end{eqnarray}     
where $u$, $v$ are real functions and 
$\nu_q$, $u_p$ are real positive wave parameters.
The latter is necessary, if we are interested in solitary waves that
exponentially decay as $|X| \rightarrow \infty$.
Introducing Eq. (\ref{35}) into Eqs. (\ref{eqs22a}-\ref{eqs22b}) we get
\begin{eqnarray}
  \label{36}
    u_{XX} - \nu^2_q u +  (u^2 + v^2) u = 0 \\
  \label{37}
    v_{XX} - \nu^2_p v +  (v^2 + u^2) v = 0 . 
\end{eqnarray}    
\begin{figure}[h]
\centerline{\hbox{
  \psfig{figure=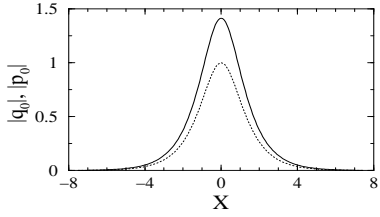,height=30mm,width=50mm,angle=-90}
}}
  \caption{
Envelopes $q_0$ and $p_0$ of the compound bright soliton
($\sigma_p = \sigma_q=+1$) for $\nu =1$ and 
$r=\Lambda_q /\Lambda_p = 2$ in arbitrary units, 
with the maximum amplitude of $p_0$ normalized to $1$.
}
\end{figure}
\noindent For $\nu_{q,p} =\nu$,  Eqs. (\ref{36}-\ref{37}) 
have a one-parameter family of symmetric and single-humped soliton
solutions (Fig. 2)
\cite{yang,jakubowski,haelterman}
\begin{eqnarray}
  \label{40}
      u(X) = \pm v(X) = \nu \, \sech(\nu X) . 
\end{eqnarray}
There are also periodic solutions of the form
\begin{eqnarray}
  \label{401}
  u(X) = A \cos(B X) \qquad 
  v(X) = A \sin(B X), 
\end{eqnarray}
where $A=\sqrt{\nu^2 + B^2 }$ with $B$ an arbitrary parameter. 
Now without loss of generality, we take $\nu_q=1$ and 
denote $\nu_p$ as $\nu$.
This can always be acheived by a rescaling of variables $u$, $v$ and $X$.
Then, for $0< \nu < 1$ there is another, in general asymmetric,
one-parameter family of solutions for each fixed $\nu$ 
\cite{yang,christodoulides}
\begin{eqnarray}
  \label{42}
    u (X) &=& { \sqrt{2(1-\nu^2)} \cosh(\nu X)}/{\kappa} \\
    \label{43}  
    v (X) &=& { -\nu \sqrt{2(1-\nu^2)} \sinh(X-X_0)}/{\kappa} ,
\end{eqnarray}
where 
\begin{eqnarray}
  \label{421}
    \kappa=\cosh(X-X_0)\cosh(\nu X)
    -\nu \sinh(X-X_0)\sinh(\nu X) \nonumber  
\end{eqnarray} 
where $X_0$ is an arbitrary parameter. 
For $X_0=0$, $u$ becomes symmetric and $v$ antisymmetric.
\begin{figure}[h]
\centerline{\hbox{
  \psfig{figure=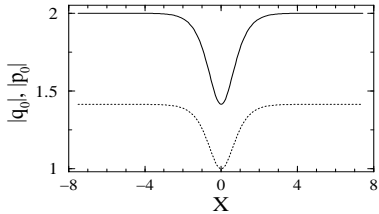,height=30mm,width=50mm,angle=-90}
}}
  \caption{
Envelopes $q_0$ and $p_0$ of the compound dark soliton
($\sigma_p = \sigma_q=-1$) for $k=1$ and 
$r=\Lambda_q /\Lambda_p=2$ in arbitrary units, 
with the minimum amplitude of $p_0$ normalized to $1$.
}
\end{figure}

For $\epsilon,\, \mu < 0$ and $\hat{a}, \hat{\beta} >0$
we have $\sigma=-1$ , and Eqs.(\ref{eqs22a}-\ref{eqs22b})
accept dark soliton solutions of the form\cite{kivsharsheppard} 
\begin{eqnarray}
  \label{51}
    Q(X,T) = P(X,T) = k \left[ \tanh(k X) - i \right] e^{i(k X - 5 k^2 T)} ,
\end{eqnarray}    
which are localized dips on a finite-amplitude background wave,
as shown in Fig. 3.
In this very interesting case of LHM the electric and 
magnetic fields are coupled together forming a dark compound
soliton.  Note that the relative amplitudes are controlled
by the corresponding nonlinearities and frequency.
For $\sigma=-1$ Eqs. (\ref{eqs22a}-\ref{eqs22b}) have
also solutions of the form\cite{kivsharsheppard}
\begin{eqnarray}
  \label{44}
    Q(X,T) = u(X) e^{-i \nu^2_q T} \, \, 
    P(X,T) = v(X) e^{-i \nu^2_p T} ,
\end{eqnarray}     
where $u$, $v$ are real functions and 
$\nu_q$, $\nu_p$  are real positive wave parameters. 
Introducing Eqs. (\ref{44}) into Eqs. (\ref{eqs22a}-\ref{eqs22b}), 
we get
\begin{eqnarray}
  \label{45}
    u_{XX} + \nu^2_q u - (u^2 + v^2) u = 0 \\
  \label{46}
    v_{XX} + \nu^2_p v - (v^2 + u^2) v = 0 . 
\end{eqnarray}    
For $\nu_{q,p} =\nu$, Eqs. (\ref{45}-\ref{46}) 
allow for kink-shaped localized soliton solutions\cite{kivshar1}
\begin{eqnarray}
  \label{49}
      u(X) = \pm v(X) = 
        ({\nu}/{\sqrt{2}}) \tanh({\nu} X /{\sqrt{2}}) ,
\end{eqnarray}
as can be  seen in Fig. 4.
In the context of the propagation of two polarization
components of a transverse EM wave in a Kerr-type medium,
the solution of this kind is often called a polarization domain wall.
There are also periodic solutions of the form
\begin{eqnarray}
  \label{50}
  u(X) = A \cos(B X) \qquad 
  v(X) = A \sin(B X), 
\end{eqnarray}
where $A= \sqrt{\nu^2 - B^2 }$ with $B<\nu$ an arbitrary parameter.  
\begin{figure}[h]
\centerline{\hbox{
  \psfig{figure=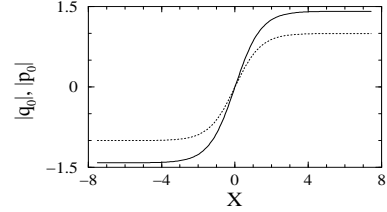,height=30mm,width=50mm,angle=-90}
}}
  \caption{
Envelopes $q_0$ and $p_0$ of the kink-shaped compound soliton
($\sigma_p = \sigma_q=-1$) for $\nu =1$ and 
$r=\Lambda_q /\Lambda_p=2$ in arbitrary units,
with the maximum amplitude of $p_0$ normalized to $1$.
}
\end{figure}

In order to check the validity of the Kerr-type approximation
for the magnetic nonlinearity,
we performed extended numerical simulations of
the complete Eqs. (\ref{weq1} - \ref{weq2})for the fields of
Eqs. (\ref{tem} -\ref{monochromatic})  using the full expression for
$\mu_{eff}$, with $\omega_{0NL}$ obtained analytically from 
Eq. (\ref{poly}), and slightly different normalization.
In the simulations we used the exact solitons of Eq. (\ref{40}) for 
right-handed medium (RHM) and Eq. (\ref{51}) for LHM
as initial
conditions and tested their subsequent time evolution and resulting stability. The
$Q$-field amplitudes are shown in Fig. 5 for a right- (two right figures) 
and a left- 
(two left figures) handed medium with $\Omega = 1.2$ and $3.0$, respectively. 
We take  the (average) amplitude of $Q$
to be unity for the right-handed medium, while the amplitude of the
background wave of $Q$ is unity for the LHM.
Similar results are obtained for the field $P$.
In the left part of Fig. 5, two bright soliton solutions 
of the form of Eq. (\ref{40}) are shown, with their difference
being the chosen value of $\nu$ in the initial conditions. 
For relative high $\nu$ the soliton is still stable,
but its amplitude is strongly oscillating.
For relatively low $\nu$ the soliton is propagating practically 
undisturbed.
In the right part of Fig. 5 two dark soliton solutions 
of the form of Eq. (\ref{51}) are shown, with their difference
being the chosen value of the constant $k$ in the initial conditions. 
Again, for relatively high $k$ the soliton developes strong osicllations 
and deforms as time progresses
while for relatively low  $k$
the soliton is propagating practically undisturbed. 
The numerics thus demonstrates clearly
that the analytical solutions of Eqs. (\ref{nls1}-\ref{nls2})  are
good approximate solutions for the complete problem at relatively small fields
while at larger amplitudes the exact solitons deform.  This is expected since magnetic nonlinearity
ceases to be of Kerr-type and saturation effects become more important.
\begin{figure}[h]
\centerline{\hbox{
  \psfig{figure=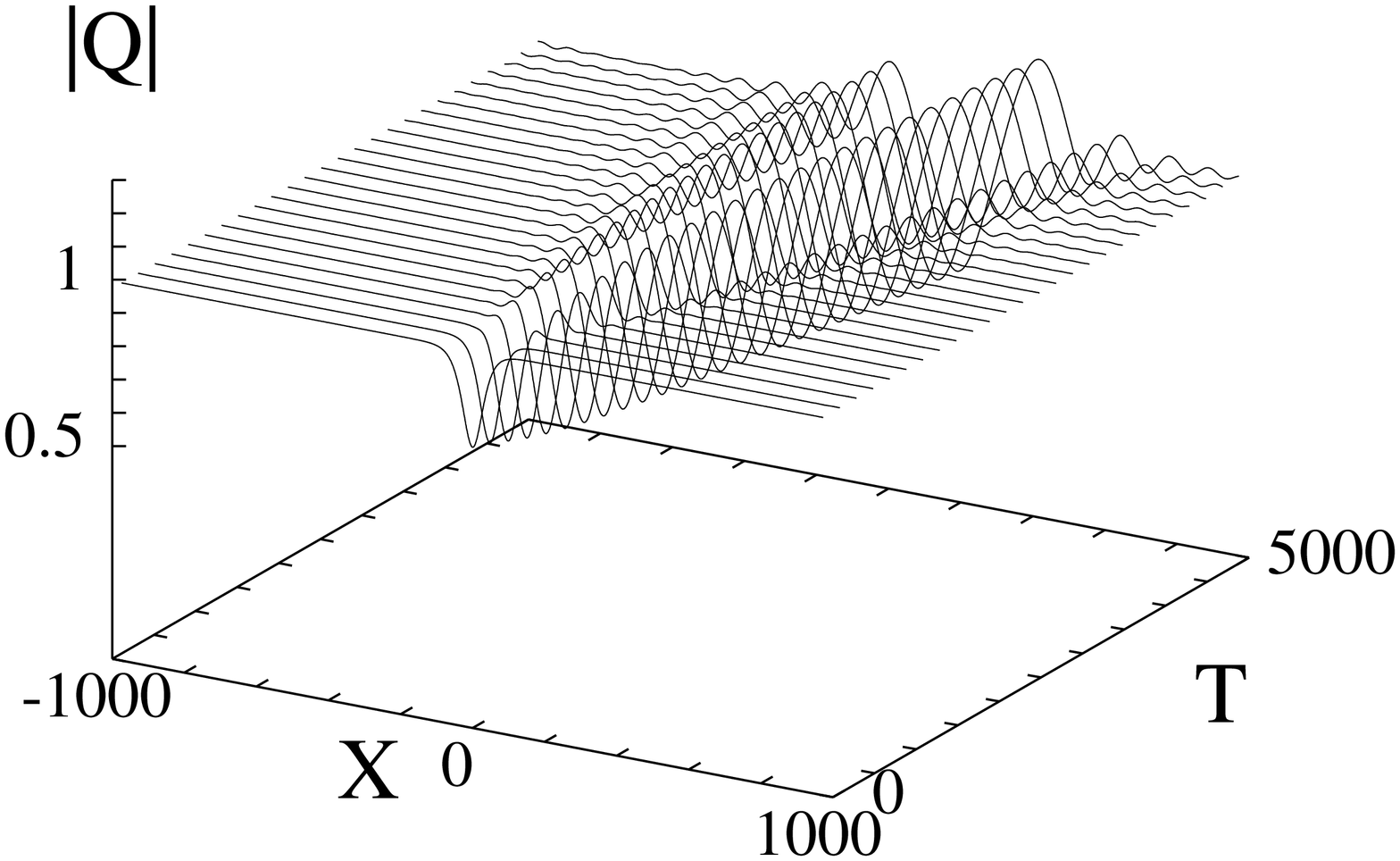,height=45mm,width=41mm,angle=-90}
  \psfig{figure=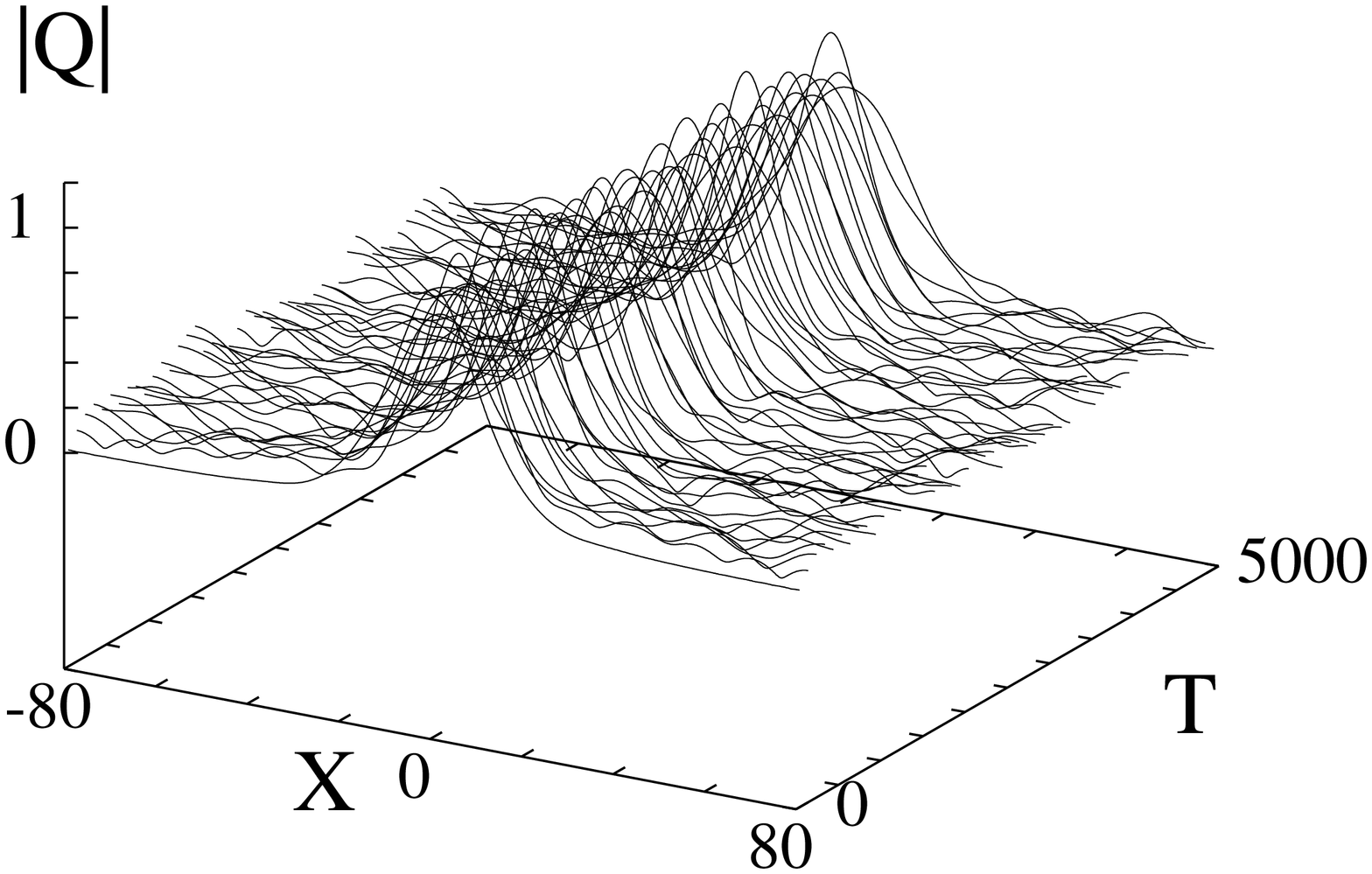,height=45mm,width=41mm,angle=-90}
}}
\centerline{\hbox{
  \psfig{figure=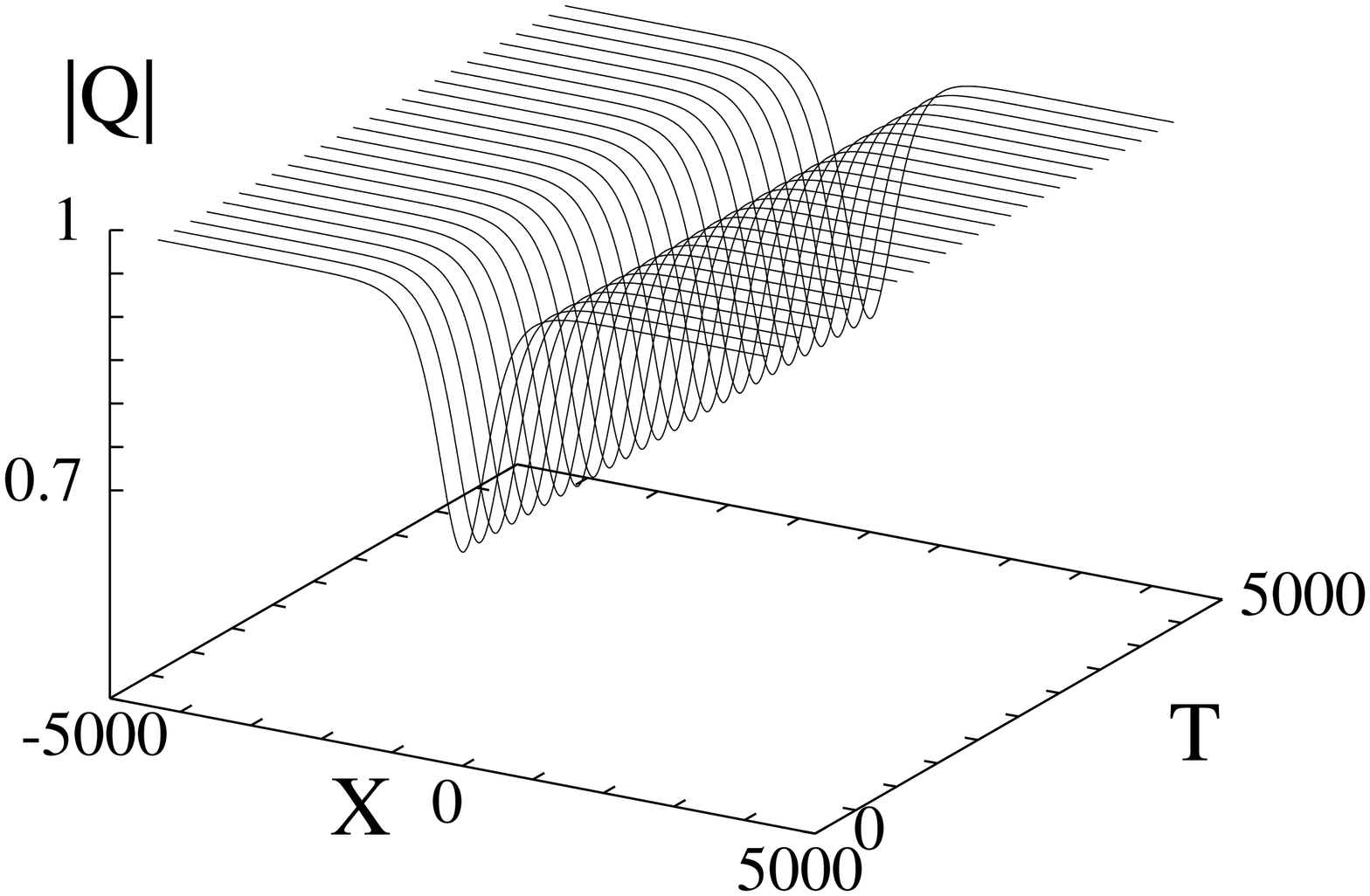,height=45mm,width=41mm,angle=-90}
  \psfig{figure=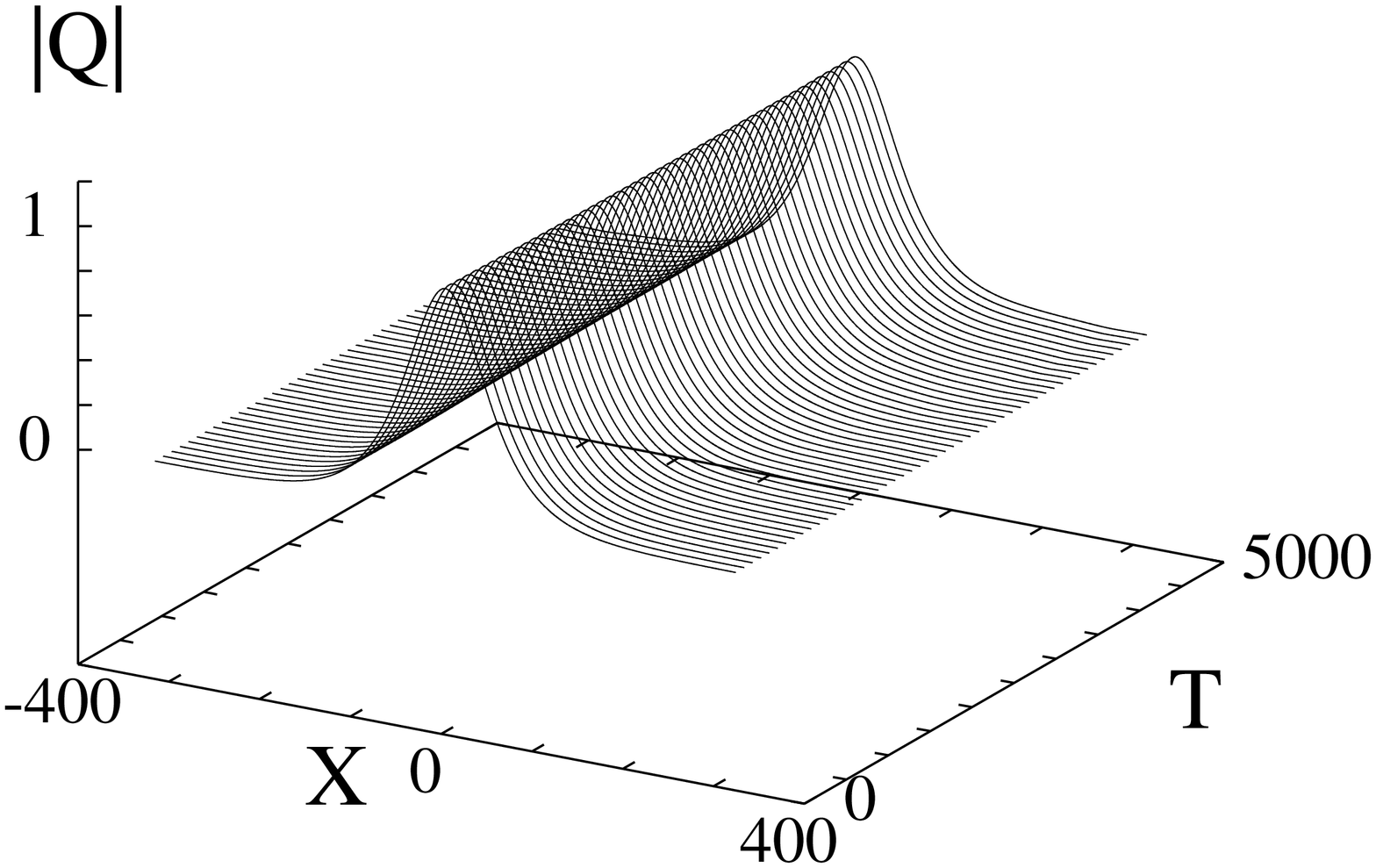,height=45mm,width=41mm,angle=-90}
}}
  \caption{Space-time evolution of solitons. 
  Right: Bright solitons (RHM) with high (upper) and 
  low (lower) amplitude $\nu$.
  Left: Dark solitons (LHM) with high (upper) and 
  low (lower) amplitude $k$. 
}
\end{figure}

In conclusion, we obtained a system of CNLS equations equivalent 
to the Manakov model describing the propagation of EM waves
in a nonlinear LHM (as well as in a nonlinear RHM) 
for relatively small fields.
Unlike a recent article,\cite{shadrivov} where only magnetic
nonlinearity was considered in the propagation of EM waves in LHMs, 
in the present work both magnetic and dielectric
nonlinearities were retained.
The present analysis however does not address the issue of the switching
effect between normal medium (RHM) and LHM properties.
Although not obvious from the Eqs. (\ref{eofomega} - \ref{poly}),
the reduction to the Manakov form becomes possible after a proper 
approximation of the complex Eqs. (\ref{mofomega}) and (\ref{poly}).
It turns out that for the choice of parameters corresponding to LHM,
the system admits compound dark soliton solutions.  
For the choice of parameters
corresponding to normal medium, the system 
admits compound bright soliton solutions.  
Reference to compound "dark" and "bright" solitons 
we mean that both soliton components, i.e. both the envelopes
of the electric and the magnetic fields, are either both dark or both
bright.  
The described effects are not limited to the specific SRR - wire  
system. They should also be present in other nonlinear LHM designs,
such as photonic crystals,\cite{markos} coupled nanowire systems,\cite{podolskiy}
the transmission line systems,\cite{eleftheriades}
and photonic systems. \cite{shvets} 
The case where the fields in the medium experience different type
of nonlinearity, leading to
$\sigma_q = -\sigma_p =1$ or $\sigma_p = -\sigma_q =1$
corresponds to a medium of positive $\epsilon_{eff}$
and negative  $\mu_{eff}$ or  negative $\epsilon_{eff}$ and
positive  $\mu_{eff}$, respectively.
This interesting case where the Manakov system does not seem to be
integrable will be treated  numerically in a following publication.


\end{document}